\documentclass[12pt]{kluwer}
\usepackage{graphicx}
\begin{document}
\begin{opening}
\title{Analysis of a Fossil Bone 
from the Archaeological Settlement
 Malu Rosu, Romania
by Accelerator Mass Spectrometry}
\author{Agata \surname{Olariu}\email{agata@ifin.nipne.ro}}
\institute{National Institute for Physics and Nuclear Engineering,
PO Box MG-6, 76900 Magurele, Bucharest, Romania}
\author{Ragnar \surname{Hellborg}\email{ragnar.hellborg@nuclear.lu.se}}
\author{Kristina \surname{Stenstr\"om}\email{kristina.stenstrom@nuclear.lu.se}}
\institute{Department of Nuclear Physics, Lund University,
S\"olvegatan 14, SE-223 62 Lund, Sweden}
\author{G\"oran \surname{Skog}\email{Goran.Skog@C14lab.lu.se}}
\institute{Department of Quaternary Geology, Lund, 
Tornav\"{a}gen 13, SE-223 63 Lund}
\author{Mikko \surname{Faarinen}\email{mikko.faarinen@nuclear.lu.se}}
\author{Per \surname{Persson}\email{per.persson@nuclear.lu.se}}
\author{Bengt \surname{Erlandsson}\email{bengt.erlandsson@nuclear.lu.se}}
\institute{Department of Nuclear Physics, Lund University}
\author{Ion V. \surname{Popescu}\email{ivpopes@ifin.nipne.ro}}
\institute{National Institute for Physics and Nuclear Engineering,
Bucharest}
\author{Emilian \surname{Alexandrescu}}
\institute{Institute of Archaeology, Bucharest}

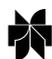
\begin{ao}\\
email: agata@ifin.nipne.ro\\
National Institute for Physics and Nuclear Engineering,\\
PO Box MG-6, 76900 Magurele, Bucharest, Romania\\
tel. 40 1 780 70 40
\end{ao}

\begin{abstract}
\noindent
~~A fossil bone from  the archaeological site Malu Rosu Giurgiu, in 
Romania has been
analyzed by accelerator mass spectrometry  to estimate its age by  
determining its
 $^{14}$C content. The radiocarbon age of the bone is in agreement
with the age obtained by the method for age determination, based on fluorine 
 content. This is  the first radiocarbon 
dating for the final Neolithic period, for this archaeological settlement in 
 the Romanian region.

\end{abstract}

\end{opening}
\section{Introduction}
~~~~Among the physical methods of importance for archaeological applications, 
 dating techniques occupy a special place. Historians need a variety of
information regarding their objects: structure, provenance, culture or
authenticity$^{1,2}$.
At the same time there is a great need of dating with precise and
objective techniques of the historically valuable material. 
There are two categories of physical methods of dating, namely one 
 in which
  the measured quantity decreases with time, and one in which the 
 physical quantity accumulates in the sample with time. 
 To the first category belong the methods in which the information on age 
 is given by the 
 disintegration rate of a radioactive nucleus, ranging from $^{40}$K, Th and U
(relatively abundant isotopes) to more rare isotopes such as
$^{14}$C, $^{10}$Be, $^{26}$Al, $^{32}$Si, $^{36}$Cl, $^{41}$Ca, $^{53}$Mn,
$^{210}$Pb.
 The second category includes the methods that are based on the  
 measurement of the accumulated 
defects produced by radiation in the environment, using 
  various techniques: thermoluminescence, electron
 paramagnetic resonance, fission traces in rocks.

Among dating methods, radiocarbon dating is the most frequently used. 
In recent decades accelerator mass spectrometry 
(AMS)$^{3,4}$, which constitutes a highly 
sensitive method for counting atoms, has been used 
for the detection of $^{14}$C.
The fact that AMS counts atoms and not decays results in great advantages
compared to radiometric techniques, requiring smaller samples and shorter
measuring times.

In the present work we have studied using the AMS technique a fossil bone 
 found in the course of the
 archaeological excavation of a workshop 
 for the processing of flintstones at
Malu Rosu, Giurgiu, near the Danube River, in Romania. 
This archaeological
assembly is believed to have an age of about 20,000 years, as based on an
historical assessment.
 Other specimens of fossil bones from Malu Rosu  have earlier been analyzed 
by a relative chemical method. The age was estimated by determining 
 the fluorine content$^{5}$ using neutron activation analysis (NAA). 
\section{Experimental}
\subsection{Preparation of the bone sample}
{\it Pretreatment of the bone sample}\\
The first step concerns the extraction of 'collagen' from the bone 
structure. Collagen is the native, biochemically intact triple-helical
macromolecule. We will use  'collagen' to refer to collagen that has undergone
a degree of diagenesis. Alterations during diagenesis are considered  to
include humification of parts of the molecule, attachment of exogenous humic
materials, and hydrolysis with loss of some amino acids.   
 We have essentially applied the Longin method$^{7}$: \\
A dried bone specimen of about 1 g is powdered using a hard knife. 
The bone powder is put into a glass vessel with a rubber cork, which is
then connected to a vacuum pump. 
About 70 - 80 ml of 2\% HCl is added in drops to
the bone powder to remove the carbonates. 
The CO$_2$ released is
 evacuated by pumping for 2 - 5 minutes. 
The bone is left in this solution for 15 - 24 h. 'Collagen' is then 
retrieved as a sediment in the vessel. 
The acid is poured off and the bone sample is washed with 
distilled water three times (until the solution is neutral).
\newpage
\noindent
The 'collagen' is transferred to a 50 ml vessel.
20 - 30 ml of distilled water is added together with 
some drops of 
2\% HCL to obtain a pH in the range 2 - 3.
This solution is kept at a temperature of 90 $^o$C 
 overnight. The pH must remain in the range 2 - 3 for the whole period of
heating, which ensures that the 'collagen' is dissolved. \\
The 'collagen' solution is filtered to remove all the
impurities, and put in the oven at 110 $^o$C 
until is 
completely  evaporated. 
The 'collagen', in gelatin form  is thus concentrated at the bottom of 
the test tube.\\
{\it Transformation to elemental carbon}\\
The next step in the preparation of the bone sample is 
 the transformation of the 'collagen'  into pure
carbon in the experimental set-up, presented in detail in refs. 6, 8 and 9. 
The conversion 
to  elemental
carbon is done in two steps: the formation of CO$_2$ by the combustion of the
 'collagen', 
 and the reduction of CO$_2$ to carbon. \\
{\small {\it Combustion}}. About 200 mg of CuO as oxidation agent is added 
to the 'collagen' sample in the test tube in order to attain 
 a 100\% combustion. 
The system is evacuated and the test tube containing the sample is
heated with a flame. The water vapor that is released during the 
combustion is caught in a cold trap.\\
{\small {\it Reduction}}. About 10 mg of metallic iron is put into a double-legged 
 test tube 
 as catalyst for
the reduction of the CO$_2$ to carbon. Prior to the reduction process, 
 the
catalyst is heated under vacuum to 650 $^o$C. With CO$_2$ transferred to and 
frozen 
in the cold trap of the
reduction part of the apparatus, hydrogen gas is
let into the system. It is suitable to have 3 times more hydrogen gas than
carbon dioxide. 
The reduction process takes about 4 h. 
When the reduction is complete, the carbon sample is dried 
by pumping for about 20 minutes.
\newpage
\noindent
The carbon sample and the iron catalyst are pressed
 into a copper  holder  with a stainless steel piston. 
 The iron catalyst provides a good
thermal conduction of the sample, which is important for the 
 sputtering process 
 into the ion source.

The copper holder with the pure carbon sample
 is put into the ion source of the accelerator. 

\subsection{The AMS analysis at the tandem accelerator}
~~~~The sample of bone, as pure carbon in the copper holder is arranged in a
wheel with 20 positions in the following sequence: 2 standards of 
oxalic acid, 1 standard of anthracite, 15 carbon samples, with the other 2
positions  being for a graphite and an optimizing sample.
Anthracite, being a very old 
coal is considered to contain no traces 
  of $^{14}$C traces and by its measurement the background for 
$^{14}$C is obtained both of the accelerator
and of the preparation procedure of the samples. Oxalic acid is a standard SRM 
prepared by 
 the USA National Bureau of Standards, with an accurately known 
 activity of $^{14}$C.
The oxalic acid is used to normalize the value of the 
$^{14}$C/$^{13}$C ratio of the 
sample. 
 The wheel with samples and standards is put into the ion source
of the accelerator.
 The central part of the Lund AMS system is a Pelletron tandem
accelerator, model 3UDH, produced by NEC, Wisconsin USA, and is 
 shown in Fig. 1. 
 The accelerator is run at 2.4 MV during AMS experiments, which is
 optimal for the C$^{3+}$ charge state when using 3 $\mu$g/cm$^2$ carbon
stripper foils. On the experimental beam line a 
 magnetic
quadrupole triplet, a velocity filter and a second analyzing magnet
  have been installed.  
The $^{13}$C$^{3+}$ beam current is usually $\approx$ 80 nA.
\newpage
\noindent
 The particle
identification and measuring system consists of a silicon surface barrier 
detector of $\Phi$=25 mm.
Part of the accelerator system is operated and controlled by a PC-based 
system. 
 The computer system analyses alternately the data of the $^{13}$C current 
received from a
current integrator and the $^{14}$C counts arriving from the particle 
detector, to obtain, finally, the ratio $^{14}$C/$^{13}$C for each sample.
 This ratio is corrected for the background (obtained from the anthracite and
oxalic acid standards) for every sample. Each sample has been measured 
 7 times. The precision of the measurements is around 1 \% .
\section{Results and Discussions}
\subsection{The radiocarbon age}
~~~~The age of the bone was determined in two steps: first the 
{\it conventional radiocarbon age} was 
calculated  and then the age was 
 converted to  calendar
 years, taking into consideration the fluctuations of the $^{14}$C 
 concentration
in the atmosphere over time.

The conventional radiocarbon age was calculated by the 
equation $^{10,11,12}$:
\begin{equation}
{\rm Age}=-8033\, \ln
\,\biggl(\frac{(^{14}C/^{13}C)_{S[-25]}}{0.9558\,(^{14}C/^{13}C)_{Ox[-19]}}\biggr)\,
{\rm years} \, {\rm BP}
\end{equation}
where $(^{14}C/^{13}C)_{S[-25]}$ is the isotopic ratio for the sample,
corrected for the background, normalized to $\delta^{13}C$=-25 $\%_0$ \\
and $(^{14}C/^{13}C)_{Ox[-19]}$ is the isotopic ratio for the oxalic acid 
 standard, corrected for the background, 
 normalized to $\delta^{13}C$=-19 $\%_0$,\\
 with $\delta^{13}C$ being:
\begin{equation}
\delta^{13}C=\biggl[\frac{R}{R_{PDB}}-1\biggr]\,\times\,1000\,\,\%_0
\end{equation}
where $R$ is the $^{13}$C/$^{12}$C isotopic ratio for the sample and the 
 oxalic acid, respectively, 
and $R_{PDB}$ is the $^{13}$C/$^{12}$C isotopic ratio of the $PDB$ 
 standard.
 The conventional radiocarbon age of the bone is expressed in years BP, 
  where present is defined to be AD 1950.
The conventional radiocarbon age for the analyzed bone from Malu Rosu has been
determined to be: 
4790$\pm$90 BP.
The conversion into calendar years was done using the OxCal 
software $^{13}$ (see Fig. 2), using data from ref. 14.
 From the processing of the data one obtained the age of the
bone from Malu Rosu to be in the interval:
\begin{center}
3760 - 3360 BC or 5710 - 5310 BP, with a confidence level of 95.4 \% . 
\end{center}
\subsection{Comparison of the radiocarbon age with the age estimated by 
 fluorine content}
~~~~The method of dating by measuring the fluorine content has been used 
 since  the end of 
the 19$^{th}$ century. The increase
of the fluorine content is the result of the contact of the fossil bone 
 with 
underground water which contains fluorine$^{15,16}$. The main inorganic 
 component of 
the bone, hydroxyapatite is slowly altered by the exchange of hydroxyl ions 
 (OH$^{-}$) with ions F$^{-}$:
\begin{center}
Ca$_{10}$(PO$_4$)$_6$(OH)$_2$ + 2F$^{-}$ $\Longrightarrow$ Ca$_{10}$(PO$_4$)$_6$F$_2$
+ 2OH$^{-}$ 
\end{center}
The result of this exchange is the formation of fluoroapatite
Ca$_{10}$(PO$_4$)$_6$F$_2$, 
a compound much more stable than hydroxiapatite.
\newpage
\noindent
The rate of accumulation of fluorine in bones depends on the environmental
conditions of the burials, the temperature and the content of fluorine in
underground waters.  \\
Some specimens of bones from Malu Rosu have been analyzed before by NAA
 to determine their fluorine content$^{5}$. 
 The irradiations were done at the
VVR-S nuclear reactor at IFIN Bucharest, at a flux 
of 10$^{12}$ neutrons/cm$^{2}\cdot $s$^{-1}$. Samples of
bones together with standards of PbF$_2$ and CaF$_2$ were irradiated 
for a period of 4 s, 
for producing
 the isotope $^{20}$F, with a half life of
 T$_{1/2}$=11 s and the main $\gamma$-ray of 1633.6 keV. The periods of
decays and times of measurement for the sample were:
t$_{decay}$=40 s, t$_{meas}$=20 s, and for the standard:
t$_{decay}$=80 s, t$_{meas}$=20 s.
For 8 specimens of bones from Romanian regions, C1, C2, C4, C5, C6, C8 and 
 C9,
 dated by radiocarbon at Groeningen,  
 the following correlation 
 was established: fluorine concentration - age (radiocarbon). For other bones:
B1 (cave bear), B2 (cave bear), B3 (cave bear), B4 (mammoth) and MR (animal
bone from Malu Rosu) was evaluated 
 the age, from the content of fluorine,
determined by NAA, using the calibration curve.
 In Fig. 3 are
presented the results of NAA of analyzed bones. Also represented are the
fluorine concentrations and the ages for some bones from Carribean 
 zone$^{13}$. In this region
 the climate is warmer and therefore the rate of accumulation of fluorine 
 in the structure
of bones is higher than in bones with the same age, from Romanian regions.
 From the diagram shown in
Fig. 3 one can estimate for
the bone from Malu Rosu, sample MR, with a content of fluorine of 400 ppm, an
 age \newline $\approx$ 6000$\pm$2000 years BP. For the bone B1 one could estimate 
 an age
around 11000$\pm$2000 years BP. For the bear B2 and the mammoth B4, and for
the bear B3, 
 one could estimate an inferior limit of 35000, respectively 37000
 years BP, 
 given by the calibration curve. 
\newpage
\noindent
Taking into consideration a possible saturation
of the curve, the bones B2 and B4, and B3 could have an age of around 40000,
respectively 45000 years BP.\\
The age of the bone from Malu Rosu of 5510$\pm$200 BP, determined by $^{14}$C 
 measurements using AMS in the present study
 is in agreement 
 with the
 estimated value of the age of 5000$\pm$1000 years BP, determined in other 
 fossil 
 bones from the Malu 
 Rosu, by measuring their content of fluorine, using NAA.
\subsection{Archaeological considerations}
~~~~The settlement at Malu Rosu Giurgiu is situated in the
south of Romania, on the bank of the River Danube, in a region where the 
 inferior
 terrace is preserved intact.

The stratigraphy of the settlement is very complex, containing, besides 
 the four
cultural Aurignacian levels (workshops for processing flintstones have been
found with some hearths in the central parts) and cultural levels from the 
 final 
 middle Neolithic (Boian culture), Eneolithic (Gumelni\c ta culture), 
la T\`{e}ne period,
early Middle Age (Dridu culture, IX-XI centuries) (Fig. 4).

Due to the specific conditions, namely open areas, the vegetal deposit and the 
loessial deposit, which start at a depth of approximately one metre, are 
affected by rodent passages both
 contemporary and fossil, 
reaching         
in some cases a depth of 6 m. Under these conditions some archaeological 
objects have been disturbed and moved from their initial position. 
 Furthermore, the Neolithic, Eneolithic and Medieval archaeological 
 complexes have 
 sometimes affected the Aurignacian cultural layers.

The fragment of bone studied by AMS in the present study originates 
 from the superior 
part of an Aurignacian layer I $^{17}$, Surface III/1995, the workshop for the 
processing of 
 flintstones Nr. 1, square D/11, depth 1.98 m (Fig. 4).
The inferior part of this Aurignacian layer, the oldest one 
 in Malu Rosu, has been dated by AMS, at Groeningen, on the basis of charcoal 
 fragments from hearths, 
 to have an 
 age of 21,140$\pm$120 BP (GrA-5094) and 22,790$\pm$130 BP (GrA-6037). 

The value of 3560$\pm$200 BC for the 
age of the bone determined by AMS in the present work  associates the bone 
with the Neolithic 
level of the Malu Rosu settlement, 
attributed by us to the final stage of evolution of the Boian culture. 
In the layer mentioned one finds, at a depth of between 
0.80 - 1.10 m, the archaeological remains of 
three houses belonging to the final stage of the Boian culture.

Some radiocarbon datings for the Boian culture can be quoted for some
settlements from the Danubian region: for the Cascioarele settlement, 
 with a layer dated
 at 4035$\pm$125 BC (BIN-800) and 3620$\pm$100 BC (BIN-796) and the settlement 
 at 
Radovanu, from the oldest layer (transitional stage to Eneolithic), dated to
3820$\pm$100 (BIN-1233)$^{18}$.
 
Thus the fossil bone analyzed in the present work, found in the
 Palaeolithic layer of the Malu Rosu settlement is dated to have an age of 
3560$\pm200$ BC, and 
belongs in fact to the
Neolithic layer, situated above the Palaeolithic one.
This result  can be explained by the existence of a dynamics of 
archaeological
objects, by which they can be moved from their original places by an 
 interference with the biosphere.  

By dating the bone from Malu Rosu, Giurgiu one obtains the first radiocarbon 
estimation of age for the final Neolithic period, for this
archaeological settlement in the Romanian region.

\newpage
\noindent
{\bf \Large References}\\ 
\noindent
1. J. W. Michels, Dating Methods in Archaeology, Seminar Press, 1973\\
2. G. A. Wagner, Archaeometric Dating, in Lectures in Isotope Geology, \\
\hspace*{0.5cm}eds. E. Jager and J. C. Hunziker, Springer, 1979, p. 178\\
3. W. Kutschera, {\it Accelerator mass spectrometry: counting atoms\\
\hspace*{0.5cm}rather than decays}, Nuclear Physics News, Vol. {\bf 3}, No. {\bf 1} (1993) 15\\
4. C. Tuniz, J. Bird, D. Fink, G. F. Herzog, Accelerator Mass \\
\hspace*{0.5cm}Spectrometry, Ultrasensitive Analysis for Global Science, CRC Press,\\
\hspace*{0.5cm}1998 \\ 
5. C. Besliu and A. Olariu, I. Popescu and T. Badica, {\it Microelements in\\ 
\hspace*{0.5cm}fossil bones and the estimation of age}, SPIE, {\bf 2339} (1995) 487, and\\
\hspace*{0.5cm}4th International Conference on Applications of Nuclear Techniques:\\
\hspace*{0.5cm}Neutrons and their Applications., Crete, Greece, 12-18 June, 1994\\
6. A. Olariu, R. Hellborg, K. Stenstr\"om, G. Skog, M. Faarinen,\\    
\hspace*{0.5cm}P. Persson, B. Erlandsson, I. V. Popescu, E. Alexandrescu,\\
\hspace*{0.5cm}{\it Analysis of a Fossil Bone from the Archaeological Settlement\\
\hspace*{0.5cm}Malu Rosu, Romania by accelerator mass spectrometry}, Report 08/00,\\
\hspace*{0.5cm}LUNDFD6/(NFFR-3081)/1-30/(2000), Lund, 2000\\
7. R. Longin, {\it New method of collagen extraction for radiocarbon dating}, \\
\hspace*{0.5cm}Nature, {\bf v 230} (1990) p. 241-242\\
8. J. S. Vogel, J. R. Southon, D. E. Nelson and T. A. Brown, Nucl. Instr.\\
\hspace*{0.5cm}and Meth. {\bf B5} (1984) p. 289\\
9. K. Stenstr\"om, G. Skog, B. Erlandsson, R. Hellborg, A. J\"anis, A.\\
\hspace*{0.5cm}Wiebert, {\it A sample preparation system for production of elemental\\
\hspace*{0.5cm}carbon for AMS analyses}, Report 02/94, LundFD6/(NFFR-3065)/\\
\hspace*{0.5cm}1-33/(1994), Lund 1994
\newpage
\noindent
10. R. Gillespie, Radiocarbon User's Handbook, Oxford University,\\    
\hspace*{0.5cm}Committee for Archaeology, Monograph Number Three, Oxonian\\
\hspace*{0.5cm}Rewley Press, 1984\\ 
11. D. J. Donahue, T. W. Linick and A. J. T. Jull, {\it Isotope ratio and\\ 
\hspace*{0.5cm}background corrections for accelerator mass spectrometry radiocarbon\\
\hspace*{0.5cm}measurements}, Radiocarbon, Vol. {\bf 32}, No. {\bf 2} (1990) p. 135\\
12. M. Stuiver, H. Polach, {\it Reporting of $^{14}$C Data}, Radiocarbon,\\
\hspace*{0.5cm}Vol. {\bf 19}, No. {\bf 3} (1977) p. 333-363\\
13. C. Ramsey, from Oxford Radiocarbon Accelerator Unit, UK, OxCal\\
\hspace*{0.5cm}software, http://www.rlaha.ox.ac.uk/oxcal/oxcal\_h.html\\ 
14. M. Stuiver, Radiocarbon, Vol. {\bf 40}, No. {\bf 3} (1988)\\
15. Z. Goffer, Archaeological Chemistry, John Wiley And Sons. Inc.\\
\hspace*{0.5cm}N.Y., 1980 \\
16. R. P. Parker and H. Toots, {\it Minor elements in fossil bones}, Geological\\
\hspace*{0.5cm}Society of American Bulletin, Vol. {\bf 81} (1970) 925\\
17. E. Alexandrescu, {\it A hypothesis about the evolution of the Aurignacian\\
\hspace*{0.5cm}cultural complex from Romanian plane} in Time of History I,
Memory \\
\hspace*{0.5cm}and Patrimonium, Bucharest University, 1997, p. 16\\
18. E. Comsa, The Neolithic on the Romanian territory - considerations,\\
\hspace*{0.5cm}Publishing House of Academy, Bucharest, Romania, 1987, p. 45.

\noindent
\vspace{18cm}

\noindent
Figure captions \\
\begin{tabular}{p{1.5cm}p{12cm}}

Fig. 1. & A schematic drawing of the Pelletron system at Lund University,
\newline for AMS studies \\

Fig. 2. & The output diagram produced by the OxCal software for the analyzed bone, showing the conversion of the conventional radiocarbon age to calendar years\\

Fig. 3. & The diagram of the fluorine concentration versus the radiocarbon age for
some fossil bones \\

Fig. 4. & The archaelogical settlement Malu Rosu Giurgiu, Romania: 
\newline a schematic drawing of the profile of the west wall of SIII, 1995
\end{tabular}
\end{document}